\documentclass{article}

\usepackage[colorlinks=true,linkcolor=blue,citecolor=blue,urlcolor=blue]{hyperref}
\usepackage[backend=biber,style=numeric,sorting=none,maxbibnames=6,url=false,eprint=false,isbn=false,doi=false]{biblatex}
\usepackage{lmodern}
\usepackage{etoolbox}
\usepackage{amsmath,amssymb,mathtools,braket}
\usepackage{siunitx}
\usepackage{overpic}
\usepackage[toc,page]{appendix}
\usepackage{authblk}
\usepackage[all]{hypcap}
\usepackage[capitalise]{cleveref}

\DeclareSIUnit\echarge{{\it e}}
\DeclareSIUnit\mbar{mbar}

\newcommand*{\sfref}[2]{Fig.~\hyperref[#1]{\ref*{#1}#2}}
\newcommand*{\sfrefs}[3]{\hyperref[#1]{Figs.~\ref*{#1}#2} and \hyperref[#1]{\ref*{#1}#3}}
\newcommand*{\calL}{\mathcal{L}}

\crefname{appendix}{Apd.}{Apds.}
\crefname{table}{Tab.}{Tabs.}
\crefname{section}{Sec.}{Secs.}

\begin{document}

\title{A Levitated Random Telegraph Noise Spectrometer}

\author[1]{Molly Message}
\author[1]{Bianca C. J. Uy}
\author[1]{Katie O'Flynn}
\author[1]{Yugang Ren}
\author[1]{Muddassar Rashid}
\author[1]{Jonathan D. Pritchett}
\author[1]{Qiongyuan Wu}
\author[2]{Hyukjoon Kwon}
\author[3]{Benjamin A. Stickler}
\author[1,4]{James Millen\thanks{Email: james.millen@kcl.ac.uk}}

\affil[1]{Department of Physics, King's College London, Strand, London, WC2R 2LS, UK}
\affil[2]{School of Computational Sciences, Korea Institute for Advanced Study (KIAS), Seoul, 02455, Korea}
\affil[3]{Institute for Complex Quantum Systems and Center for Integrated Quantum Science and Technology, Ulm University, Albert-Einstein-Allee 11, Ulm, 89069, Germany}
\affil[4]{London Centre for Nanotechnology, Department of Physics, King's College London, Strand, London, WC2R 2LS, UK}

\date{}

\maketitle

\begin{abstract}
\mathversion{normal} Random Telegraph Noise is a ubiquitous process manifesting across technology and the natural world. It is characterized by random jumps between two distinct states with Poissonian waiting times, and is the origin of $1/f$ noise. Understanding and characterizing this noise is critical for the reliable operation of micro-, nano- and quantum-technologies. In this work we probe random telegraph noise using a levitated microparticle sensor whose dynamics are driven almost entirely by this non-white source of noise. We observe a startling resonant behaviour, characterized by a thousand-fold increase in the underdamped sensor's position fluctuations, enabling us to measure the spectral properties of the noise over six decades of timescale. This work not only provides a unique way to probe random telegraph noise, but also demonstrates a platform for studying non-equilibrium stochastic dynamics in the presence of realistic non-white noise, with applications from biology to social behaviour.
\end{abstract}

\noindent\textbf{Keywords:} Levitated electro-dynamics, Random telegraph noise, Sensing

\section*{Introduction}\label{sec1}
There is no such thing in nature as white noise, despite its ubiquity in theoretical models and colloquial terminology. Real systems operate in environments where structured and coloured noise plays a decisive role in how the system evolves \cite{PhysRevB.91.174102,GREENHALGH2016684,PhysRevE.85.031106}, motivating investigations into stochastic dynamics beyond the white-noise approximation. However, coloured noise introduces memory effects that often render the dynamics analytically intractable \cite{Haunggi_1994}. Tractable coloured-noise models therefore provide a powerful framework for studying realistic stochastic systems. Among the most physically-relevant examples is Random Telegraph Noise (RTN), a two-level stochastic process for which switching rates are governed by Poisson jumps.

Since its early observation in vacuum and solid-state electronic devices \cite{Simoen_2011}, RTN arising from the trapping and release of charge carriers at material defects \cite{Martin-Martinez2020} has been investigated for its impact on voltage and current fluctuations. It is well established that the cumulative effect of many RTN sources constitutes a microscopic origin of low-frequency $1/f$ noise in electronic systems \cite{PhysRevLett.119.097201}. As electrical components are miniaturized to the nanoscale, RTN signals from only a few, or even a single, defect can significantly influence the operation of transistors, capacitors, memory devices and quantum dots \cite{Simoen_2011,Martin-Martinez2020}. Analysis of RTN has therefore become an important tool for detecting and characterizing material defects \cite{Rev.Sci.Instrum.71,9926003,10.1063/5.0295457}. Beyond electronics, RTN has been used to model ATP-driven dynamics and epidemiological processes \cite{PhysRevE.93.052604,GREENHALGH2016684}, as well as financial stock price fluctuations \cite{Ratanov2008TelegraphMarkets}.

RTN is a non-Gaussian coloured-noise process characterized by a single parameter, its switching rate $\nu$. Studying the influence of RTN within a Brownian harmonic oscillator framework provides a simple and tractable platform for exploring coloured-noise-driven dynamics, with previous theoretical work demonstrating strongly enhanced escape rates \cite{PhysRevE.85.031106} and the generation of directed currents from zero-mean asymmetric RTN driving of a potential barrier height \cite{Barik_2006}. Despite its widespread use in theoretical studies, experimental investigations of RTN-driven dynamics remain limited, partly because experimental platforms for stochastic systems are typically dominated by pseudo-white noise environments \cite{Millen2018Review,Franosch2011}.

Levitated particles are a paradigmatic system for studying stochastic dynamics \cite{G-Ballestero2021rev,MillenReview2020, Seifert_2025}, enabling foundational investigations of phenomena such as ballistic Brownian motion \cite{Li2010} and the Kramers process \cite{Rondin2017}. While optical tweezers have dominated in this field, Paul traps have been used to realise a single-ion heat engine \cite{Rossnagel2016} and to operate a heat engine at temperatures exceeding $10^7$\,K \cite{Message_2025}. Their deep trapping potentials ($>10^9$\,K) and absence of optical absorption enable stable trapping of strongly driven underdamped systems over extended periods.

We levitate a charged silica microsphere in a Paul trap and study its motion when coupled to a synthetic RTN bath across a wide range of switching rates, $\nu$. We conduct our experiment in the underdamped regime, preventing the RTN-driven motion from fully dissipating between switching events. This gives rise to a resonance \cite{Militaru2021EscapeActiveParticles,Nozaki1999ColoredNoiseSR} as $\nu$ approaches the levitated oscillator's motional frequency, $\omega_z$. The deep Paul trap potential maintains stable harmonic confinement even during resonance. We develop analytical descriptions in both the time and frequency domains and show that the RTN characteristics become encoded in the particle motion. Inspired by levitated sensors \cite{Liang2023Yoctonewton}, we use the levitated particle as a probe of the coloured noise across six orders of magnitude of switching rate.

\section*{A charged microparticle levitated in a Random Telegraph Noise bath}\label{sec2}

\begin{figure}[ht]
\centering
\begin{overpic}[width=0.99\linewidth]{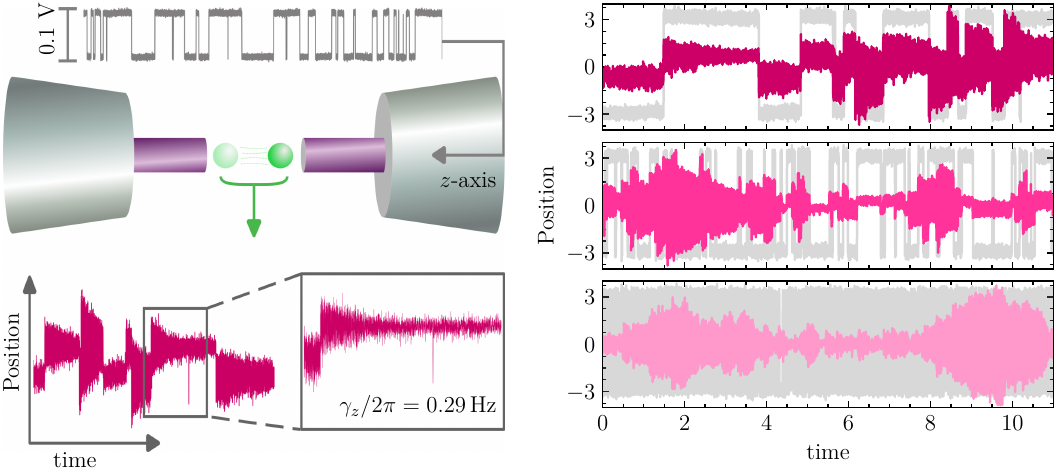}
    \put(0,42){a)}
    \put(50,42){b)}
\end{overpic}
\caption{\textbf{A levitated charged microparticle coupled to an RTN bath} a) RTN is synthesized as a voltage (gray trace), which is amplified and applied to an electrode near the levitated particle, producing displacements along the $z$-axis (magenta trace). The inset shows a zoom-in around a single RTN jump, with a displacement followed by ringdown at the damping rate $\gamma_z$. b) Three examples of the particle's 1-D motion under different RTN rates $\nu$ of $1\,s^{-1}$ (top panel), $10\,s^{-1}$ (middle panel) and $10^2\,s^{-1}$ (bottom panel). The position signals in each trace are scaled by the position standard deviation $\sigma_z$ across the trajectories. The applied RTN is shown on each sub-plot in grey, with amplitude 0.1\,V.}
\label{fig:experiment_illustration}
\end{figure}

We levitate a silica microsphere of diameter $(4.82\pm0.50)\,\unit{\um}$ and charge $(16490\pm180)\unit{\echarge}$ in a linear Paul trap. The trap comprises four electrodes arranged at the corners of a square, generating a quadrupole potential that confines the particle in the radial plane. Confinement along the trap axis, defined here as the $z$-direction, is provided by two coaxial control electrodes positioned on either side of the trapping region (see \sfref{fig:experiment_illustration}{a}), which are also used to apply electrical forces to the levitated particle. Further details of the experimental system are given in \cref{apd:sec:experiment}. The motion of the levitated particle is tracked with high spatial and temporal resolution using an event-based camera (EBC)~\cite{Ren_2022}.

The trapped particle behaves as a three-dimensional harmonic oscillator with centre-of-mass mode frequencies $
\omega_x/2\pi= (78.80\pm0.03)\,\unit{\Hz}$, $\omega_y/2\pi=(172.80\pm0.02)\,\unit{\Hz}$ and $\omega_z/2\pi=(275.75\pm0.01)\,\unit{\Hz}$. In this harmonic potential, the secular motion is decoupled along each axis and remains so even at temperatures exceeding $10^7\,$K \cite{Message_2025}.
The system is maintained at a pressure of $P=(2.3\pm0.4) \times 10^{-3}\,\unit{\mbar}$, where collisions with residual gas provide a momentum damping rate $\gamma_z/2\pi = (0.29\pm0.03)\,\unit{\Hz}$, placing the system firmly in the underdamped regime $\omega_{x,y,z} \gg \gamma_z$. In this regime, the thermal forces from the surrounding gas is weak compared to the force we exert on the particle by applying electric fields. Although the experiment operates at room temperature, voltage fluctuations in the trapping electronics raise the effective equilibrium temperature to $T_z=(530\pm 50)\,\unit{\K}$.

Applied electrical fields can be used to synthesize a thermodynamic environment for charged micro-objects, for example mimicking pseudo-white noise baths \cite{Message_2025,Martinez2016}. Here, we extend this approach to generate a coloured noise environment, in this case binary RTN. The RTN is synthesized (National Instruments FlexRio FPGA) and applied as a voltage to one of the control electrodes $d = 0.83\,\unit{\mm}$ away from the levitated particle (see \sfref{fig:experiment_illustration}{a}). It switches between two discrete states $\eta_t=\{+1,-1\}$ with amplitude $\Delta V=0.1\,\unit{\V}$ at a characteristic rate $\nu=1/\tau$, where $\tau$ is the mean waiting time between jumps. The signal is generated by first selecting an initial state $\eta_0$ at random and holding it for a duration $t_\tau$ before switching. The dwell time $t_\tau$ is sampled from
\begin{equation}\label{equ:sample_telegraph_noise}
    t_\tau=-\,\mathrm{ln}(\alpha)/\nu,
\end{equation}
where $\alpha$ is a random variable uniformly distributed between $0$ and $1$. Subsequent dwell times are generated in the same way throughout the experiment, yielding Poisson-distributed switching events (see \cref{apd:sec:telegraph}). 

The applied RTN generates a one dimensional force along the $z$-direction, influencing the motion of the levitated microparticle. In \sfref{fig:experiment_illustration}{b}, example trajectories of the levitated particle are shown for three RTN switching rates: $\nu = 1\,s^{-1}$ (top panel), $10\,s^{-1}$ (middle panel) and $10^2\,s^{-1}$ (bottom panel), corresponding to slow, intermediate and fast driving relative to the particle's oscillation frequency $\omega_z/2\pi$. For slow switching, the RTN signal is clearly resolved in the particle motion. At higher switching rates, the dynamics become qualitatively similar to those of a system driven by white noise.

\section*{The response of the levitated particle to RTN}\label{sec3}

\begin{figure*}[t]
    \centering
    \begin{overpic}[width=0.99\linewidth]{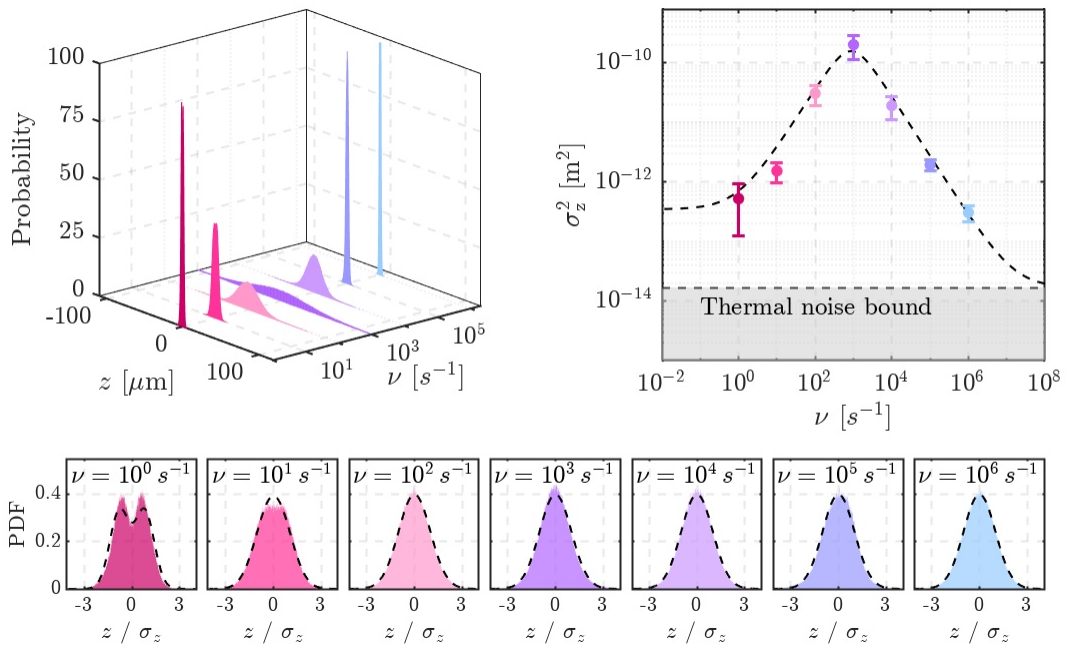}
    \put(0,59){a)}
    \put(53,59){b)}
    \put(0,18){c)}
    \end{overpic}
    \caption{\textbf{Position probability density functions (PDFs) of a particle driven by RTN} a) The particle's position PDF at different RTN switching rates $\nu$. The $z$-axis is truncated for clarity. b) Variation in position variance $\sigma_z^2$ with $\nu$ (coloured points), extracted from the time-series of the data and compared to the theoretical model in \cref{equ:position_variance} with no free parameters (black dashed line). c) Individual position PDFs for different $\nu$, with the position axis normalized by the standard deviation presented in b), compared to the theoretical model in \cref{equ:position_distributions} with no free parameters.
    }
    \label{fig:VariancePlot2}
\end{figure*}

The motion of the levitated microparticle along the $z$-axis can be modelled by a stochastic Langevin equation subject to both white noise and RTN, such that
\begin{equation}\label{equ:system_langevin}
    \ddot{z}_t = -\gamma_z \dot{z}_t-\omega_z^2 z_t + \sqrt{\frac{2\gamma_z k_BT_z}{m}} \zeta_t + \frac{qE_z}{m} \eta_t,
\end{equation}
where $z_t$ is the particle's centre-of-mass position at time $t$, $\gamma_z$ is the gas damping rate, $\omega_z$ the oscillation frequency, $m$ the particle mass, $T_z$ the temperature of the white-noise environment, $q$ the particle charge and $E_z = \Delta V / d$ the electric field generated along the $z$-axis by applying a voltage difference $\Delta V$ across the control electrodes. Thermal fluctuations are represented by a normalised white noise process $\zeta_t$ satisfying $\mathbb{E}[\zeta_t] = 0$ and $\mathbb{E}[\zeta_t \zeta_{t'}]=\delta(t-t')$, while the RTN term $\eta_t$ satisfies $\eta_t=\{-1,+1\}$, $\mathbb{E}[\eta_t] = 0$ and $\mathbb{E}[\eta_t \eta_{t'}]=\exp({-2\nu\lvert t-t'\rvert})$ (see \cref{apd:sec:telegraph}). 

The characteristic energy scale associated with the white and random telegraph noise are $k_BT_z$ and $(qE_z)^2/m\omega_z^2$, respectively. Considering the experimental parameters used in this study, we find the RTN-to-white-noise energy ratio as $17\gg 1$, indicating that the dynamics are dominated by the telegraph noise.

\textit{Position probability distribution functions --}
The influence of the RTN switching rate $\nu$ on the dynamics of the levitated particle is first studied via the position probability distribution functions (PDFs) shown in \sfref{fig:VariancePlot2}{a}. A resonance-like enhancement of the position fluctuations occurs as the switching rate approaches half the levitated particle's oscillation (angular) frequency, with an increase in variance of three-orders-of-magnitude at $\nu=10^3\,s^{-1}$ relative to $\nu=1\,s^{-1}$ and $\nu=10^6\,s^{-1}$. High resolution particle tracking across this wide range is enabled by imaging with an EBC \cite{Ren2024,Message_2025}. 

To characterise the motion of the levitated particle, we derive the moments from  \cref{equ:system_langevin} in \cref{apd:sec:theory}. The position variance under underdamped conditions is given by,
\begin{equation}\label{equ:position_variance}
    \sigma_z^2 = \frac{k_B T_z}{m\omega_z^2} + \frac{q^2E_z^2}{m^2\omega_z^4}\frac{1}{1 +(2\nu/\omega_z)^2} ( 1+2\nu/{\gamma_z}),
\end{equation}
where the RTN-induced variance increases as $\gamma_z$ decreases, reflecting that weaker damping enhances the response of the system to the RTN. The experimentally measured position variance is presented in \sfref{fig:VariancePlot2}{b}, showing good agreement with \cref{equ:position_variance}. The variance exhibits a maximum at $\omega_z/\nu=2$, a resonance-like response driven by the RTN.

We observe a qualitative transformation of the position PDF between the slow-switching limit $\nu \lesssim \gamma_z$ and the fast-switching limit $\nu \gg \gamma_z$, as shown in \sfref{fig:VariancePlot2}{c}.
In the slow-switching limit, where the system equilibrates between switching events, the motion follows a bi-modal Gaussian profile with each peak centred at $\pm qE_z/m\omega_z^2$. In the fast switching limit, where the RTN is well approximated as white noise $\mathbb{E}[\eta_t \eta_{t'}] \approx \delta(t - t') / \nu$, the motion reduces to a single Gaussian distribution. Since no simple closed-form expression for the spatial distribution exists in the intermediate regime, we approximate the position PDF as,
\begin{subequations}\label[equation]{equ:position_distributions}
    \begin{align}
        P(z) &= \frac{1}{2}\left [ \mathcal{N}(z-\mu,\sigma) + \mathcal{N}(z+\mu,\sigma) \right ],\\
        \text{with} \qquad \mu &= \frac{qE_z}{m\omega_z^2}\frac{1}{1+(2\nu/\omega_z)^2}, \qquad \sigma^2 = \sigma_z^2-\mu^2.
    \end{align}
\end{subequations}
This expression recovers the bimodal and Gaussian limits as $\nu\to0$ and $\nu\to\infty$, respectively. The transition from a bimodal distribution to a single Gaussian occurs at $\nu\sim10\,s^{-1}$ for our parameters, where higher even standardized cumulants become negligible (see \cref{apd:sec:theory}). As can be seen in \sfref{fig:experiment_illustration}{b}, the influence of the discrete RTN jumps can only be seen in the time-series of the particle's motion in the very slow switching limit. Hence, we need to look beyond the position statistics to use the levitated particle as a probe of the RTN. 

\begin{figure}[t]
    \centering
    \begin{overpic}[width=\linewidth]{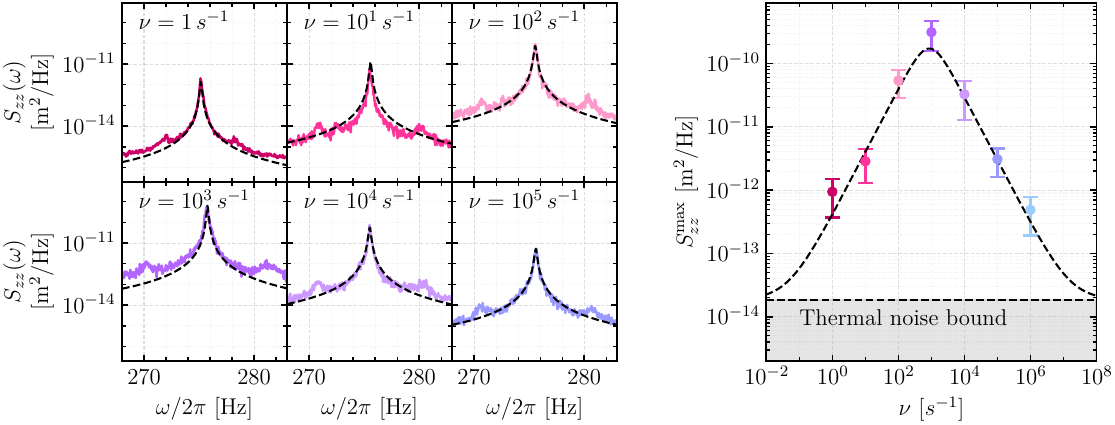}
        \put(0, 36){a)}
        \put(60.5,36){b)}
    \end{overpic}
    \caption{\textbf{Spectral analysis of a particle driven by RTN} a) Position PSDs of the particle's motion at different RTN characteristic rates $\nu$. The data (coloured solid lines) is compared to the theoretical model in \cref{equ:psd_equation} with no free parameters (black dashed lines). b) The maximum of the position PSDs, extracted by analysing a 0.2\,Hz window around $\omega = \omega_z$ (coloured points), compared to the theoretical prediction in \cref{equ:maximum_PSD} with no free parameters (black dashed line). In this figure, the data for each value of $\nu$ is averaged over $10$ time traces, each with a duration of $30$\,\unit{\s}. 
    }
    \label{fig:PSDPlot}
\end{figure}

\textit{Power spectral density --} Analysis of the position power spectral density (PSD) provides a complementary approach for characterising the influence of RTN on the dynamics of our levitated particle sensor. The PSD of the particle motion under both white noise and RTN can be derived from \cref{equ:system_langevin}, and is given by
\begin{align}\label{equ:psd_equation}
    S_{zz}(\omega) =\left(\frac{2\gamma_zk_BT_z}{m}+ \frac{q^2E_z^2}{m^2}\frac{4\nu/\omega^2}{1+(2\nu/\omega)^2}\right)\frac{1}{(\omega^2-\omega_z^2)^2+\gamma_z^2\omega^2}.
\end{align} 
As shown in \sfref{fig:PSDPlot}{a}, the experimental PSDs agree well with this model. From \cref{equ:psd_equation}, the maximal size of the PSD (at $\omega=\omega_z$ for underdamped dynamics) is given by, 
\begin{equation}\label{equ:maximum_PSD}
    S_{zz}^{\max}(\nu)=\frac{2}{\gamma_z}\left( \frac{k_BT_z}{m\omega_z^2} + \frac{q^2E_z^2}{m^2\omega_z^4}\frac{2\nu/\gamma_z}{1+(2\nu/\omega_z)^2}\right).
\end{equation}
We compare the maxima of the experimental PSDs with this model in \sfref{fig:PSDPlot}{b}, finding excellent agreement. Note that the maxima of the PSDs exhibit a resonance-like effect with varying RTN switching rate $\nu$, with the peak response at $\omega_z/\nu=2$. This enhancement is larger for smaller damping rates. Derivations of Eqs.~\ref{equ:psd_equation} and \ref{equ:maximum_PSD} are provided in \cref{apd:sec:theory}.

Given the agreement between our presented models and experiments, and the absence of free parameters in  \cref{equ:position_variance,equ:psd_equation,equ:maximum_PSD} (all parameters can be independently measured when there is no applied noise), we demonstrate a protocol for sensing and characterizing RTN using a levitated particle sensor.

\section*{Sensing the spectral characteristics of RTN}

We present a sensing protocol reliant upon the resonance-like dependence of the levitated particle's position PSD on the RTN switching rate $\nu$, as seen in \cref{fig:PSDPlot}. We extract the value of $\nu$ by fitting the experimentally measured PSD with \cref{equ:psd_equation}, where the sharp spectral response of an underdamped oscillator facilitates a precise extraction of the peak height. Due to the fact that, for a fixed particle frequency $\omega_z$, the shape of each experimental PSD generally corresponds to one of two possible values of $\nu$ (which cannot be discerned with a single measurement), we introduce a protocol which uses frequency-ramping of the levitated sensor's resonant frequency $\omega_z$ to uniquely sense the true value of the applied RTN switching rate $\nu$. 

\begin{figure}[tb]
    \centering
    \begin{overpic}[width=\linewidth]{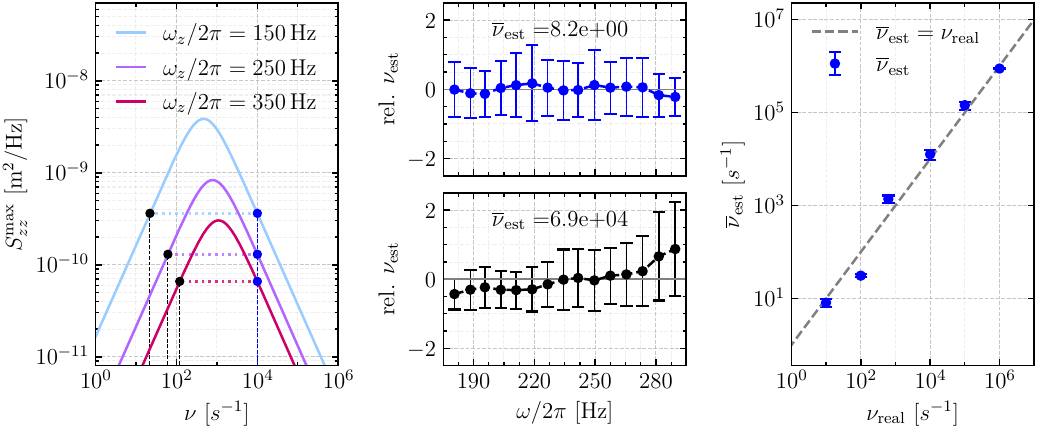}
        \put(0, 40){a)}
        \put(35.5,40){b)}
        \put(68.2,40){c)}
    \end{overpic}
    \caption{\textbf{Sensing the spectral characteristics of RTN with a levitated sensor} a) The theoretical values of $S_{zz}^{\max}(\nu)$ for three values of $\omega_z$ (solid lines). Each value of $S_{zz}^{\max}$ corresponds to two possible values of $\nu$ (blue and black points). The true value of $\nu$ does not vary with $\omega_z$ (blue points). b) Experimental data from running the sensing protocol and extracting two values of $\nu_{\mathrm{est}}$ for each value of $\omega_z$ (blue and black points). The black points vary with $\omega_z$, whereas the blue points fluctuate around their mean value, indicating the correct value of $\nu_{\mathrm{est}}$. c) The comparison between $\nu_{\mathrm{est}}$ and $\nu_{\mathrm{real}}$ using our sensing protocol, showing good agreement. Each value of $\nu_{\mathrm{est}}$ is extracted from at least 10 values of $\omega_z$.}
    \label{fig:Sensitivity}
\end{figure}

The working principle of the sensing protocol with frequency-ramping is illustrated in \sfref{fig:Sensitivity}{a}, which shows three curves of $S_{zz}^{\max}$ against $\nu$ (from \cref{equ:maximum_PSD}) at different particle frequencies $\omega_z$. At each frequency $\omega_z$, sensing a true RTN switching rate (for example, $\nu_\mathrm{real} = 10^{4}\,\unit{\s^{-1}}$) yields two estimates $\nu_\mathrm{est}$ from fitting the experimental PSD with \cref{equ:psd_equation}, which cannot be distinguished. However, as the frequency $\omega_z$ increases, one of these estimates remains fixed while the
other increases with $\omega_z$. The fixed value of $\nu_\mathrm{est}$ corresponds to the true RTN switching rate $\nu_\mathrm{real}$.

We experimentally demonstrate this method by applying RTN with $\nu_\mathrm{real}=10^{1}\,\unit{\s^{-1}}$ and sweeping $\omega_z$ from $2\pi\times(180\to 290)\,$Hz by changing the voltage used to levitate the particle. The relative variation of the two sets of fitted $\nu_\mathrm{est}$ is shown in \sfref{fig:Sensitivity}{b}, where we define
\begin{equation}
    \mathrm{rel.~} \nu_\mathrm{est} = \frac{\nu_\mathrm{est} - \overline{\nu}_\mathrm{est}}{\overline{\nu}_\mathrm{est}},
\end{equation}
with $\overline{\nu}_\mathrm{est}$ the mean of each set. Clearly, one set of estimates fluctuates around its mean value (blue) while the other increases with $\omega_z$ (black). The estimated RTN rate in this example is $\overline{\nu}_\mathrm{est} = (8.2 \pm 0.9) \,\unit{\s^{-1}}$.

In \sfref{fig:Sensitivity}{c} we demonstrate that this method can be used to sense $\nu_\mathrm{real}$ over 6 orders of magnitude, provided that the RTN amplitude exceeds the thermal noise bound as shown in \sfref{fig:PSDPlot}{b}. Further details are provided in \cref{apd:sec:sensing_data}. The points at $\nu_{\mathrm real} = 10^2\,$s$^{-1}$ and $10^3\,$s$^{-1}$ provide the worst estimates, which we attribute to the interaction with the harmonic of line noise ($100\,$Hz) and the on-resonance interaction between the sensor and the RTN, respectively.

Our presented sensing protocol offers two advantages. First, it allows us to distinguish the background RTN from white noise, by examining the frequency-dependence of the estimated $\nu_\mathrm{est}$ (i.e.~whether $\nu_\mathrm{est}$ increases with the the particle's frequency $\omega_z$, as shown by the black dots in \sfref{fig:Sensitivity}{b}). Secondly, this protocol allows us to sense the spectral characteristics of RTN even when $\nu$ is so small that the noise cannot be resolved in the temporal response of the sensor. This protocol requires knowledge of the electric field $E_z$, which must be independently measured. 
In principle, one can estimate $\nu$ using the resonance relation $\omega_z/\nu=2$. In this case the knowledge of $E_z$ is not required, but the sensor would have to be tuned over the desired frequency range and comes with a wider uncertainty (see \cref{apd:sec:resonance_sensing}).

\section*{Discussion}
In this work, we measure the position PDF and PSD of an electrically levitated microsphere driven by RTN across a wide range of switching rates, $\nu$. We develop analytic theoretical models of the coloured dynamics, which show excellent agreement with experimental data. We observe a resonance-like particle response when the switching rate approaches the particle's oscillation frequency, and present a method for extracting the underlying noise switching rate from the resulting dynamics.

The induced resonant response, together with previous observations of modified escape rates and stochastic resonance in coloured noise driven systems \cite{PhysRevE.85.031106,Militaru2021EscapeActiveParticles,Nozaki1999ColoredNoiseSR}, point to a more general consequence of coloured noise in stochastic dynamics: its ability to enhance or suppress system behaviour. This is particularly relevant in biological systems, where resonance phenomena and escape dynamics underpin a range of processes such as neuronal signal transduction \cite{Nozaki1999ColoredNoiseSR}. Levitated particle sensors therefore offer a controllable physical platform for the analogue simulation of stochastic dynamics in coloured-noise environments, enabling future investigations into how structured noise influences biologically relevant non-equilibrium processes. 

When considering the ability to probe the noisy environment, previous measurements of RTN signals from material defects have been taken using the time-resolved response of a sensor, via, for example, frequency-modulated atomic force microscopy \cite{Cowie_2024}. Levitated sensors offer a complementary approach as their operating frequency is widely tuneable, set by the trapping potential rather than material characteristics. As a result, a levitated particle system can be configured to probe RTN across a wide spectral range. The exceptional force sensitivity achieved by levitated sensors \cite{Liang2023Yoctonewton} also opens the possibility of detecting extremely weak RTN sources. Finally, the use of spectral analysis makes the system robust against challenges typically associated with time domain analysis, such as background noise obscuring the underlying RTN signal.

\section*{Acknowledgements}
JM notes primary support for this project from The Leverhulme Trust under Leverhulme Research Grant RPG-2024-017. JM also recognises funding from the European Research Council (ERC) under the European Union’s Horizon 2020 research and innovation programme (grant agreement no. 803277) and from the Engineering and Physical Sciences Research Council (EP/S004777/1). BAS is supported by the DFG through grant No.
510794108 and by the Carl-Zeiss-Foundation through the Carl-Zeiss Stiftungscenter QPhoton.  H.K. is supported by the KIAS Individual Grant No. CG085302 at Korea Institute for Advanced Study and by the National Research Foundation of Korea
(NRF) grant funded by the Korea government (MSIT) (RS-2024-00413957 and
RS-2024-00438415).

\subsubsection*{Supplementary information}
Provided as Appendices for now.

\subsubsection*{Author contributions}
MM designed and ran the experiment, analysed data and wrote the manuscript. BU acquired data. JP analysed data and helped prepare the manuscript. KOF built the experiment and acquired data. YR assisted in running the experiment. MR designed and built the experiment. QU analysed data, built theoretical models and wrote the manuscript. HK and BS built the theoretical models and helped design the experiment. JM conceived of and designed the experiment, and helped prepare the manuscript.  

\newpage
\begin{appendices}
\crefalias{section}{appendix}

\section{Experimental details}\label{apd:sec:experiment}
A $(4.8\pm0.5)\,\mu$m diameter silica sphere (Bangs Laboratories, Inc.) is levitated at $(2.0\pm0.4)\times10^{-3}$\,mbar using a linear Paul trap. The Paul trap consists of four 3.0\,mm diameter steel rods that are positioned at the four corners of a square, with the centres of the rods on a circle of radius 5.0\,mm. A signal generator (Stanford Research Systems DS345) generates a sinusoidally varying voltage which is amplified (TREK 10/10B-HS) and applied to one pair of diagonally opposed electrodes as V~=~$V_{\rm{RF}}\rm{cos}(\omega_{\rm{RF}}t)$. Each newly trapped particle has a different charge-to-mass ratio and therefore requires different trap parameters for stable confinement. The particle specific parameter pairs ($V_{\rm{RF}}$, $\omega_{\rm{RF}}$) used in this work were (4000~V, $2\pi\times 1100$~Hz),(4200~V, $2\pi\times 920$~Hz),(X~V, $2\pi\times Y$~Hz),(X~V, $2\pi\times Y$~Hz). These four electrodes generate a time-average harmonic potential in the $y$-$z$ plane. The other pair of diagonally opposed electrodes are used to position the particle at the centre of the potential to minimize its micromotion, and carry 0-10\,V DC.

Two 1.0\,mm diameter steel rods (endcaps) are aligned coaxially along the center of the Paul trap with a separation of 1.66\,mm. A voltage supply (Stanford Research Systems SIM928) generates a DC voltage that is amplified (Falco Systems WMA-20) to give $U_0$ = -8.0\,V on both electrodes, confining the particle in 3D.

The particle is trapped using Laser Induced Acoustic Desorption \cite{Bykov2019,Nikkhou2021} at a pressure of $4\times10^{-2}\,$mbar. A dry sample of microparticles is sonicated for 30 minutes, and subsequently spread onto a 0.4\,mm thick aluminum sheet. A second sheet of aluminum is placed on top and rubbed across the sample, positively charging the particles in excess of $10^4\,e$. This method and the mass-selectivity of the Paul trap leads to trapping of single spheres, as confirmed by light scattering.
A 532 nm laser beam (Vortran Stradus) of 25\,mW power and a beam waist radius of $\sim100\,\mu$m is used to image the particle, which scatters light onto the sensor of an Event Based Camera (EBC), (Prophessee EVK3 Gen4.1), which tracks the particle in real time. This allows us to track the particle over hundreds of micrometers while retaining a position resolution of $30\,$nm\,Hz$^{-1/2}$.
Calibration of our imaging system, calculation of particle charge and use of the EBC is described in detail in ref. \cite{Ren2022}. Once the system has been calibrated, the temperature of the particle can be calculated by analyzing the power spectral density or the position variance using \cite{MillenReview2020}
\begin{equation}
    S_{zz}^\mathrm{cali}(f) = 2\pi\cdot\frac{2\gamma_0 k_B T_0/m_0}{((2\pi f)^2-\omega_0^2)^2 + \gamma_0^2(2\pi f)^2} + c,
\end{equation}
which gives us the fittings and uncertainties in the main text.

Using the LabVIEW FPGA module (National Instruments), the two-level telegraph noise with adjustable amplitude, switching rate and DC offset, is generated and converted to an analogue voltage by an NI FPGA R-Series XYZ. The telegraph noise is then added to a single endcap electrode.

\section{Modelling telegraph noise}\label{apd:sec:telegraph}
\begin{figure}[t]
    \centering
    \includegraphics[width=0.45\textwidth]{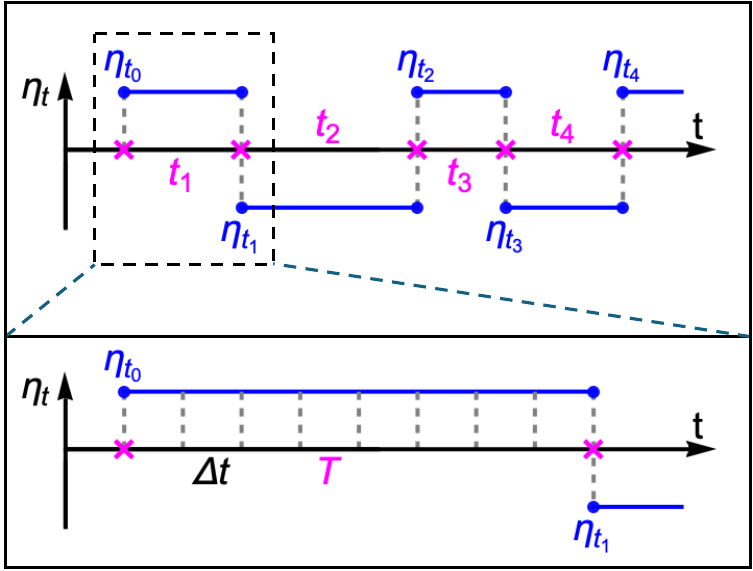}
    \caption{Illustration of one Telegraph jump and its connection to the Poisson process.}
    \label{fig:apd:Telegraph_noise_illustrate}
\end{figure}

We model the binary telegraph noise with a random variable $\eta_t\in\{+1,-1\}$ where the switches of its value follow the Poisson process with a constant rate $\nu=1/\tau$, such that the number of jumps $n$ in time $t$ follows the Poisson distribution,
\begin{equation}
    P^{\mathrm P}_{n,t} = \frac{\left(\nu t\right)^n}{n!} e^{-\nu t}.
\end{equation}
The transition probabilities can be calculated,
\begin{subequations}
    \begin{alignat}{2}
        P(\eta_t=\pm1 \vert \eta_0=\pm1) &=\sum_{k=0}^\infty P^{\mathrm P}_{2k,t} &= \frac{1}{2} + \frac{1}{2}e^{-2\nu t}, \\
        P(\eta_t=\pm1 \vert \eta_0=\mp1) &=\sum_{k=0}^\infty P^{\mathrm P}_{2k+1,t} &= \frac{1}{2} - \frac{1}{2}e^{-2\nu t},
    \end{alignat}
\end{subequations}
from which we find that the telegraph process $\eta_t$ is Markovian and stationary, \textit{i.e.}~$P(\eta_t\vert\eta_{t'},\eta_0) = P(\eta_t\vert\eta_{t'})$, $P(\eta_t\vert\eta_0) = P(\eta_0\vert\eta_t)$, $P(\eta_t)= \frac{1}{2}$, and the correlation in time reads
\begin{align}
    \mathbb{E}[\eta_t \eta_{t'}] &= \sum_{\eta_t, \eta_{t'}\in\pm 1}\eta_t\eta_{t'}P(\eta_t\vert\eta_{t'})P(\eta_{t'}), \notag\\ 
    &=e^{-2\nu\lvert{t-t'}\rvert }.
\end{align}
To efficiently sample the telegraph noise consisting of a number of jumps $\{\eta_{t_n}\}$, one needs to find the probability density for the waiting time $T$ between jumps. As shown in \cref{fig:apd:Telegraph_noise_illustrate}, the probability of having one jump (and $N$ no jumps) in a time interval $T = N \Delta t$ reads
\begin{align}
    \mathrm{Prob}(T) &= \left(\frac{1}{2} - \frac{1}{2}e^{-2 \nu\Delta t}\right)\left(\frac{1}{2} + \frac{1}{2}e^{-2 \nu\Delta t}\right)^N, \notag \\
    &\approx\nu \Delta t \left(1-\frac{\nu T}{N}\right)^N, \notag \\
    &\approx\nu\Delta t \,e^{-\nu t},
\end{align}
where we take the first-order approximation on the exponential. Then, the probability density of the waiting time between two jumps is given by $P(T)= \mathrm{Prob}(T) / \Delta t$, which yields
\begin{equation}\label{apd:equ:telegraph_time}
    P(T) = \nu e^{-\nu T},
\end{equation}
with $\int_0^\infty P(T)\,\mathrm{d}T =1$ and the mean value $\overline{T} =1/\nu = \tau$. 

In the experiment, we sample the the time between two jumps with \cref{equ:sample_telegraph_noise}, whose characteristic function reads
\begin{align}
    \mathbb{E}[e^{i k t_\tau}] &= \int_{0}^{1}\mathrm{d}\alpha \,e^{-ik \mathrm{ln}(\alpha)/\nu}, \notag \\
    &=\frac{\nu}{\nu - i k}.
\end{align} 
The corresponding probability distribution of $t_\tau$, $P(t_\tau) = \frac{1}{2\pi}\int_{-\infty}^{\infty}\mathrm{d}k~e^{-ik t_\tau}~\mathbb{E}[e^{i k t_\tau}]$, gives \cref{apd:equ:telegraph_time}, indicating that sampling the time $t_\tau$ with \cref{equ:sample_telegraph_noise} generates a telegraph voltage signal with Poisson distributed jumps.

\section{Cumulants of the motion}\label{apd:sec:theory}
Here we compute the cumulants of the particle's motion under white and telegraph noises in the long time limit $t\gg 1/\gamma_z$, where the dependence on the initial condition can be neglected. The cumulants can be computed through the stationary moment hierarchy. We define the infinitesimal generator $\calL$ acting on the function $f(z_t,v_t,\eta_t)$ from the dynamics in \cref{equ:system_langevin} as
\begin{align}
\calL f & =\lim_{\mathrm{d}t\to0}\frac{\mathbb{E}\left[f(z_{t+\mathrm{d}t},v_{t+\mathrm{d}t},\eta_{t+\mathrm{d}t})\mid z_t,v_t,\eta_t\right]-f(z_t,v_t,\eta_t)}{\mathrm{d}t}, \notag\\
&= v_t \frac{\partial f}{\partial z_t} +(-\gamma_z v_t -\omega_z^2 z_t + A\eta_t)\frac{\partial f}{\partial v_t}  + \frac{\sigma^2}{2}\frac{\partial^2 f}{\partial v_t^2} + \nu\left[f(z_t,v_t,-\eta_t)-f(z_t,v_t,\eta_t)\right],
\end{align}
where we define $v_t = \dot{z}_t$, $A = q E_z/m$ and $\sigma = \sqrt{2\gamma_zk_BT_z/m}$. The It\^o's formula is used for describing the white-noise contribution $\sigma^2$, and the jump probability for $\eta_{t+\mathrm{dt}}$ is $\nu \mathrm{d}t$. The stationary moment hierarchy is the expectation of the generator $\mathbb{E}[\calL f]=0$ with $f= z_t^iv_t^j\eta_t^r$, which reads
\begin{equation}\label{apd:equ:moment_generator}
    i M_{i-1,j+1}^r + j\left[-\gamma_z M_{i,j}^r - \omega_z^2M_{i+1,j-1}^r + A M_{i,j-1}^{1-r}\right] + \frac{\sigma^2}{2} j (j-1) M_{i,j-2}^r - 2\nu r M_{i,j}^r = 0,
\end{equation}
where $M_{i,j}^r = \mathbb{E}[z_t^iv_t^j\eta_t^r]$ and $r=0,1$ as $\eta_t^{2n} = 1$ and $\eta_t^{2n+1} = \eta_t$.

The moments of the motion can be calculated with \cref{apd:equ:moment_generator}. The first moments can be attained by
\begin{equation}
    M_{0,0}^1=\mathbb{E}[\eta_t]= 0, \qquad M_{0,1}^0=\mathbb{E}[v_t]=0, \qquad M_{1,0}^0=\mathbb{E}[z_t]=0,
\end{equation}
and the same for the second moments,
\begin{align}
    &M_{1,0}^1=\mathbb{E}[z_t\eta_t] = \frac{A}{\omega_z^2} \frac{1}{1+(2\nu/\omega_z)^2 + 2\gamma_z\nu/\omega_z^2}, \quad M_{0,1}^1=\mathbb{E}[v_t\eta_t] = 2\nu \mathbb{E}[z_t\eta_t], \notag \\
    &M_{2,0}^0=\mathbb{E}[z_t^2] = \frac{\sigma^2}{2\gamma_z\omega_z^2} + \frac{A^2}{\omega_z^4} \frac{1+2\nu/\gamma_z}{1+(2\nu/\omega_z)^2 + 2\gamma_z\nu/\omega_z^2}, \quad M_{1,1}^0=\mathbb{E}[z_tv_t]=0, \notag\\
    &M_{0,2}^0=\mathbb{E}[v_t^2] =\frac{\sigma^2}{2\gamma_z} + \frac{A^2}{\omega_z^2} \frac{2\nu/\gamma_z}{1+(2\nu/\omega_z)^2 + 2\gamma_z\nu/\omega_z^2},
\end{align}
which gives \cref{equ:position_variance} in the main text under the assumption $\gamma_z\ll\omega_z$.

\begin{figure}[t]
    \centering
    \includegraphics[width=0.5\textwidth]{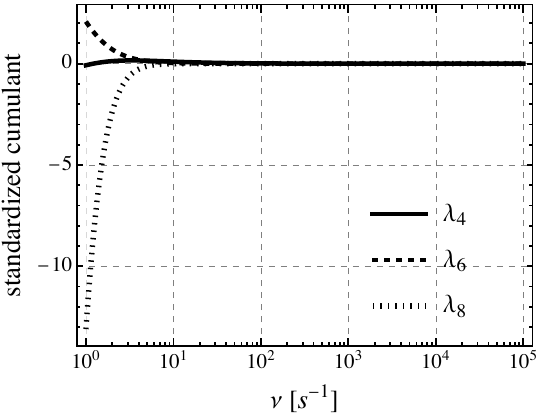}
    \caption{Numerically computed standardized cumulants against the RTN switch rate $\nu$.}
    \label{fig:apd:cumulants}
\end{figure}

Higher moments can be obtained by solving the corresponding \cref{apd:equ:moment_generator} as demonstrated above. Their expressions can be efficiently attained using symbolic software, such as Mathematica. As an example, the excess kurtosis, describing the fourth-order deviation from the Gaussian distribution, reads
\begin{align}
\lambda_4
&=
\frac{\mathbb{E}[z_t^4]-3\mathbb{E}[z_t^2]^2}
{\mathbb{E}[z_t^2]^2}
=
\frac{
24 A^4 \gamma_z \omega_z^2\,\mathcal P_4
}{
\mathcal D_4
},
\end{align}
where
\begin{align}
\mathcal P_4
={}&
-6(\gamma_z+\nu)(\gamma_z+2\nu)^3(3\gamma_z+2\nu)
-(\gamma_z+2\nu)
(33\gamma_z^2+172\gamma_z\nu+104\nu^2)\omega_z^2
\notag\\
&+6(-2\gamma_z+3\nu)\omega_z^4,
\\[1ex]
\mathcal D_4
={}&
\left[2(\gamma_z+\nu)(\gamma_z+2\nu)+\omega_z^2\right]
\left(3\gamma_z^2+4\omega_z^2\right)
\left(6\gamma_z\nu+4\nu^2+9\omega_z^2\right)
\notag\\
&\times
\left[
2A^2(\gamma_z+2\nu)
+
\left(2\nu(\gamma_z+2\nu)+\omega_z^2\right)\sigma^2
\right]^2 .
\end{align}
In \cref{fig:apd:cumulants}, we show evaluations of higher-order standardized cumulants $\lambda_4$, $\lambda_6$, and $\lambda_8$. It can be seen that their values approach $0$ in the fast-switching limit $\nu\gg \gamma_z$.

For the odd moments, given the unbiased initial telegraph noise state $\mathbb{E}[\eta_0]=0$, the position distribution of the particle motion $z_t$ is symmetric in distribution, $P(z_t) = P(-z_t)$. The odd moments thus read
\begin{equation}
    \mathbb{E}[z_t^{2n+1}] = \int_{-\infty}^\infty \mathrm{d}z_t\, z_t ^{2n+1} P\left(z_t\right) = 0.
\end{equation}
Namely, all odd moments vanish due to the symmetry of the dynamics.

\textit{PSD of particle's motion --}
Take the Fourier transformation $\tilde{z}_\omega = \frac{1}{\sqrt{t}}\int_0^t\mathrm{d}t' e^{i\omega t'}z_{t'}$ and use the relation $\dot{z}_t = i\omega \tilde{z}_\omega$, the particle's motion by \cref{equ:system_langevin} can be described in frequency space such that
\begin{equation}
    -\omega^2 \tilde{z}_\omega  = i\omega\gamma_0\,\tilde{v}_\omega-\omega_0^2 \tilde{z}_\omega + \zeta_0\tilde{\zeta}_\omega+\eta_0\tilde{\eta}_\omega,
\end{equation}
which gives particles' motion in the frequency domain with
\begin{equation}\label{equ:motion_in_frequency_domain}
    \tilde{z}_\omega = \frac{1}{\omega_0^2-\omega^2+i\gamma_0\omega}\left(\zeta_0\tilde{\zeta}_\omega + \eta_0\tilde{\eta}_\omega \right),
\end{equation}
where $\tilde{z}_\omega$, $\tilde{\zeta}_\omega$ and $\tilde{\eta}_\omega$ are the Fourier transformations of particle's motion $z_t$ and noises terms $\zeta_t$ and $\eta_t$. The PSD can then be acquired by 
\begin{align}\label{apd:equ:psd_equation}
    S_{zz}(\omega) &= \lim_{t\to\infty}\mathbb{E}[z_\omega z_\omega^\ast] \notag\\
    &=\lim_{t\to\infty}\frac{\zeta_0^2\mathbb{E}[\zeta_\omega \zeta_\omega^\ast]+\eta_0^2\mathbb{E}[\eta_\omega \eta_\omega^\ast]}{(\omega^2-\omega_0^2)^2+\gamma_0^2\omega^2}.
\end{align}
From the definition, $\mathbb{E}[\zeta_\omega \zeta_\omega^\ast] =1$ while
\begin{align}
    \lim_{t\to\infty} \mathbb{E}[\eta_\omega \eta_\omega^\ast]&= \lim_{t\to\infty} \frac{1}{t}\iint_0^t\mathrm{d}t_1\mathrm{d}t_2\,e^{i\omega(t_1-t_2)}\mathbb{E}[\eta_{t_1} \eta_{t_2}] \notag\\
    &=\frac{\tau_0}{1+\tau_0^2\omega^2/4},
\end{align}
with $\mathbb{E}[\eta_{t_1} \eta_{t_2}]=\exp({-2\lvert t_1-t_2\rvert/\tau_0})$. Put them back to \cref{apd:equ:psd_equation} gives \cref{equ:psd_equation} in the main text.

\section{Details on the sensing protocols}\label{apd:sec:sensing_data}
In general, each sensing trail takes more than $10$ experimental runs, where the particle's oscillation frequency ramps over a range exceeding $100\,\unit{\Hz}$. In this study, each experimental run recodes the particle motion for $30\,\unit{\s}$ at a sampling rate $1\,\unit{\kHz}$. The recorded data is converted to physical displacement with a pre-measured calibration factor. The particle charge and the electric-field strength are independently measured before the trail. 

The PSD of the motion data is then computed. The particle's oscillation frequency, damping rate and RTN switching rate are extracted by fitting the PDF to the theoretical model given by \cref{equ:psd_equation} using the least square method. The fitting is performed over a frequency window of approximately $10\,\unit{\Hz}$ around the particle's oscillation frequency. We found that using fewer PSD points improves the sensing performance, which is achieved by segments when computing the experimental PSD. The fit also returns parameter uncertainties, which are propagated when estimating the final error range.

For each experimental run, the RTN-rate sensing can yield two possible solutions. This occurs because the system have similar responses to each solution, as shown by \sfref{fig:Sensitivity}{a}. Both solutions are recorded together with the corresponding mechanical frequency.
After collecting all outcomes in the sensing trail, we plot the estimated RTN rates as a function of the mechanical frequency. From the theoretical model, the counterfeit estimation can be identified as the curve with the larger positive gradient $k$. Therefore, the code automatically selects the data set corresponding to the curve with the smaller gradient. The selected data set is then used to compute the mean RTN-rate estimate and its associated error range, which are reported in \sfref{fig:Sensitivity}{c}. 

Estimating the precision of our sensing protocol is challenging due to the span over many orders of magnitude of frequency. Instead, we compare $\nu_{\mathrm est} = 10^{\tilde{n}}\,$s$^{-1}$ to $\nu_{\mathrm{real}} = 10^n\,$s$^{-1}$ for the data in \sfref{fig:Sensitivity}{c}. In this case, we find a reduced $\chi^2$ value of 0.42 in the comparison of $n$ to $\tilde{n}$. Finally, we found that this sensing technique behaves better at lower pressure because high damping weakens the RTN-induced response of the system and limits the range of sensibility.

\section{Sensing using the resonance relation}\label{apd:sec:resonance_sensing}

\begin{figure}[tb]
    \centering
    \begin{overpic}[width=\linewidth]{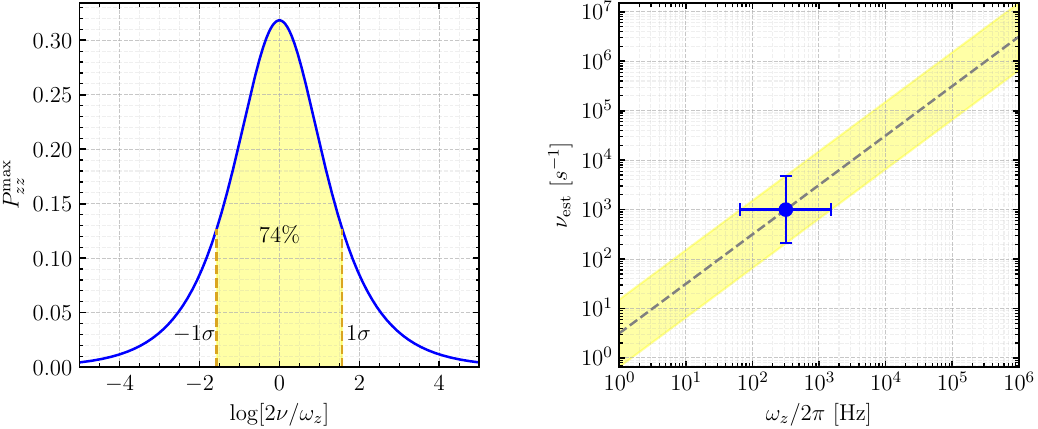}
        \put(0, 40){a)}
        \put(52,40){b)}
    \end{overpic}
    \caption{Analysis of sensing using the resonance relation. Panel a) shows the probability distribution $P_{zz}^{\max}$ against the logarithmic dimensionless variable $\log\left[2\nu/\omega_z\right]$. Panel b) shows the required particle ramping frequency (horizontal bar) to measure the RTN characteristic rate $\nu_\mathrm{est}$ with one standard deviation (vertical bar).}
    \label{apd:fig:Sensitivity}
\end{figure}

To demonstrate the sensing technique using the resonance relation $\omega_z/\nu=2$, transform the maximum PSD by \cref{equ:maximum_PSD} into a probability distribution by introducing a dimensionless variable $y \equiv \log(2\nu/\omega_z)\in[-\infty, \infty]$, such that
\begin{align}
    P_{zz}^{\max}(y) &= \frac{2}{\pi}\frac{\gamma_z}{\omega_z}\frac{m^2\omega_z^4}{q^2E_z^2} \left(\frac{\gamma_z}{2}\,S_{zz}^{\max}(\nu) - \frac{k_BT_z}{m\omega_z^2} \right),  \notag \\
    &=\frac{2}{\pi} \frac{e^y}{1 + e^{2y}}, \notag\\
    &=\frac{1}{\pi}\mathrm{sech}(y)\label{equ:sensing_prob}
\end{align}
and $\int_{-\infty}^{\infty} P_{zz}^{\max}(y)\,\mathrm{d}y = 1$. This probability distribution has the mean $\mu=0$ (corresponding to the estimate $\nu_\mathrm{est}=\omega_z/2$) and the standard deviation $\sigma = \pi/2$, as shown in the \sfref{apd:fig:Sensitivity}{a}. This indicates that, roughly, this technique produces an estimation with a wide uncertainty $\nu_\mathrm{est}\in\left[e^{-\pi/2}\omega_z/2, e^{\pi/2}\omega_z/2\right]$. Assuming that faithful sensing requires the particle ramping frequency to cover one standard deviation in this probability distribution, this corresponds to ramping over more than 1 order-of-magnitudes in frequency. For the example shown in \sfref{apd:fig:Sensitivity}{b}, if the RTN rate to be sensed is $\nu_\mathrm{real}=10^{3}\,\unit{\s^{-1}}$, the particle's frequency $\omega_z/2\pi$ needs to ramp from $66\,\unit{\Hz}$ to $1531\,\unit{\Hz}$, and the estimated RTN rate has a one-standard-deviation uncertainty range of $207\,\unit{\s^{-1}}\leq\nu_\mathrm{est}\leq4810\,\unit{\s^{-1}}$.

\end{appendices}

\printbibliography

\end{document}